# Electronic structure and chemical bonding of nc-TiC/a-C nanocomposites

Martin Magnuson[1], Erik Lewin[2], Lars Hultman[1] and Ulf Jansson[2]

[1]*Department of Physics, Chemistry and Biology (IFM), Linköping University, SE-58183 Linköping, Sweden.*

[2]*Department of Materials Chemistry, The Ångström Laboratory, Uppsala University, P.O. Box 538 SE-75121 Uppsala*

**Abstract**

The electronic structure of nanocrystalline (nc-) TiC/amorphous C nanocomposites has been investigated by soft x-ray absorption and emission spectroscopy. The measured spectra at the Ti 2*p* and C 1*s* thresholds of the nanocomposites are compared to those of Ti metal and amorphous C. The corresponding intensities of the electronic states for the valence and conduction bands in the nanocomposites are shown to strongly depend on the TiC carbide grain size. An increased charge-transfer between the Ti 3$d$-$e_g$ states and the C 2*p* states has been identified as the grain size decreases, causing an increased ionicity of the TiC nanocrystallites. It is suggested that the charge-transfer occurs at the interface between the nanocrystalline TiC and the amorphous C matrix and represents an interface bonding which may be essential for the understanding of the properties of nc-TiC/amorphous C and similar nanocomposites.

# 1 Introduction

Nanocomposites comprise materials with two or more phases for which at least one has nanometer-size crystallites [1]. Carbon-based nanocomposites of nanocrystalline metal-carbides (nc-MC) embedded in an amorphous carbon (a-C) matrix, are materials with inherent design possibilities for applications utilizing mechanical, tribological, and/or electrical properties. By tuning the TiC grain size and fraction of a-C matrix, the properties can be controlled as the matrix phase increases the toughness of the material and softens the nanocomposite compared to pure carbides but also provides a source of solid lubricant [2,3,4,5,6,7,12]. The nc-TiC has a NaCl crystal structure and its bonding is a mixture of covalent, ionic and metallic bonds where the covalent contribution consists of Ti $e_g$ - C 2*p* ($pd_\sigma$), Ti $t_{2g}$ - C 2*p* ($pd_\pi$), and Ti - Ti $t_{2g}$ ($dd_\sigma$) bonds [9], which are all found in the valence band.

    Previous experimental investigations of the electronic structure of nc-TiC/a-C nanocomposites have used core-level x-ray photoelectron spectroscopy (XPS) [10,11,12,13]. Core-level C 1*s* XPS measurements of TiC are known to exhibit two different spectral components originating from C-C and C-Ti bonding. The spectral component of the C-Ti bonding has a rather large (~3 eV) chemical shift towards lower binding energy in comparison to the C-C component in a-C, which is a signature of electronic charge transfer from Ti to C. For the Ti 2*p* core levels, the

chemical shift of the XPS peaks of TiC is small compared to the large shift at the C 1*s* core level. An interesting observation is that the C 1*s* XPS spectrum of the nc-TiC/a-C nanocomposite is different compared to single phase TiC. In the nanocomposite C 1*s* spectrum, an additional shoulder is observed at about ~ 283 eV binding energy between the C-C and Ti-C peaks, which we denote C-Ti$^*$ [11,14,3,7,15,16,17,18]. The intensity of this additional feature increases upon sputter-etching [19]. However, recent studies with high energy XPS have shown that it is present also in unsputtered samples. Furthermore, the relative intensity of the C-Ti$^*$ feature increases with reduced TiC grain size [11]. The fact that it is almost only observed in nanocomposites and related to the grain size suggest that it can be attributed to an interfacial state at the TiC-matrix interface. The presence of such interfacial state of varying quantity may strongly affect the physical, chemical and mechanical properties of the nanocomposites. For a-C films, the relative intensity of the ~ 283 eV feature is also known to depend on the differences in sample structure or composition caused by different deposition methods [20,21].

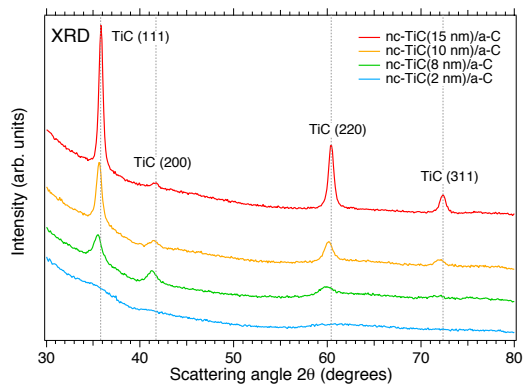

**Figure 1:** (Color Online) XRD from the nc-TiC/a-C films with different grain size (in parenthesis).

Additional information about the unoccupied states can be achieved with soft x-ray absorption (SXA) spectroscopy in either surface sensitive total electron yield (TEY) mode or bulk-sensitive fluorescence yield (TFY) mode. A previous SXA investigation of C:V, C:Co, C:Cu nanocomposite films grown by ion beam cosputtering indicate that the interfaces are indeed important to take into account [22]. The C 1*s* x-ray absorption spectra of carbon materials usually exhibit a rather sharp and characteristic $\pi^*$ absorption peak and a broad well-separated $\sigma^*$ shape resonance at more than 5 eV higher energy. However, the origin of a C 1*s* absorption feature located between the $\pi^*$ and $\sigma^*$ absorption peaks in transition metal carbide materials has been controversial [23,24] and may be due to C-O bonding, $sp^2$-$sp^3$ hybridization or hybridized C 2*p* - transition metal 3*d* - 4*sp* states at the interfaces [22].

The aim of this study is to increase the understanding of the nature of the electronic structure, the interface state and chemical bonding in nc-TiC/a-C nanocomposites using SXA spectroscopy in TEY and TFY modes in combination with soft x-ray emission (SXE) spectroscopy. The SXA technique probes the unoccupied electronic states, while the SXE technique probes the occupied electronic states in the materials. For probing the occupied states, the SXE technique is more bulk sensitive than electron-based spectroscopic techniques which is useful when investigating internal and embedded electronic structures and interfaces. The combination of SXA and SXE measurements on nc-TiC/a-C nanocomposites with different carbide grain sizes (i.e. varying interface/bulk ratio) gives valuable information on possible intermediate interface states. In addition, the different types of C contributions are selected by tuning the excitation energies for the SXE measurements which give additional insight into the controversial nature of the ~ 283 eV absorption peak feature between the $\pi^*$ and $\sigma^*$ C 1s absorption resonances.

## 2 Experimental

### 2.1 Deposition of nanocrystalline films

The deposition of the nc-TiC/a-C and a-C films were carried out in an ultra high vacuum chamber (base pressure 10 Torr) by non-reactive, unbalanced DC-magnetron sputtering from separate 2 in. elemental targets, supplied by Kurt J. Lesker Company Ltd. (purity specified as 99,995% and 99,999% for Ti and C, respectively). Through tuning of magnetron currents, samples with a carbon content between 35 and 100 at. % were deposited. The sample thicknesses were kept constant at about 0.2 μm. The Ar-plasma was generated at a constant pressure of 3.0 mTorr, with a flow rate of Ar into the chamber of 150 sccm. Substrates were placed about 15 cm below the magnetrons on a rotating substrate holder. Deposition was carried out simultaneously on amorphous SiO and Si(111)-substrates. The films deposited on SiO-substrates were used in the SXA/SXE experiments and XRD, while the films deposited on Si-substrates were used for XPS analysis. The substrates were heated to 300C by a BN-plate with W-wires, situated about 5 mm below the substrates. The temperature was monitored using a Mikron M90-0 infrared pyrometer, calibrated against a TiC thin film using a thermo-element. Prior to deposition, the substrates were preheated for 30 minutes, and the targets were presputtered for 10 minutes.

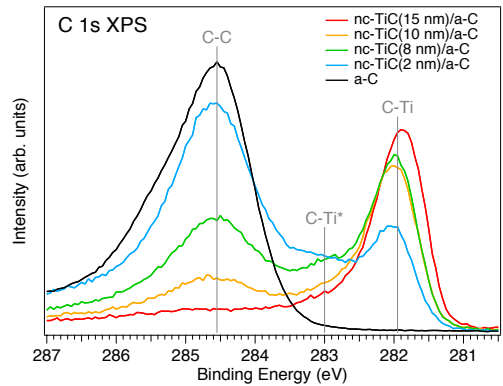

**Figure 2:** (Color Online) A series of C 1s XPS spectra of the nc-TiC/a-C films with different grain size (in parenthesis), plotted on a binding energy scale. The spectra were measured after sputter-etching.

The samples were analyzed with X-ray diffraction (XRD) using a Philips MRD X'pert diffractometer with parallel beam geometry and grazing incidence (GI-XRD) scans using a 2 incident angle. XPS was performed using a Physical Electronics Quantum 2000 ESCA Microprobe. The chemical composition (C and Ti content) was determined from the XPS sputter depth profiles, using sensitivity factors calibrated against a bulk TiC reference sample to calculate the composition with regards to Ti and C. The C bonding was analysed by high resolution XPS acquisitions after sputter etching to a depth of 150 Å. Etching was performed with low ion energy (200 V Ar) to minimise sputter damage; ion beam was rastered over a 1x1 mm area, and the analysis spot was 200 μm in diameter.

## 2.2 X-ray absorption and emission measurements

The SXA and SXE measurements were performed at the undulator beamline I511-3 at MAX II (MAX-lab National Laboratory, Lund University, Sweden), comprising a 49-pole undulator and a modified SX-700 plane grating monochromator [25]. The SXE spectra were measured with a high-resolution Rowland-mount grazing-incidence grating spectrometer [26] with a two-dimensional detector. The Ti *L* and C *K* SXE spectra were recorded using a spherical grating with 1200 lines/mm of 5 m radius in the first order of diffraction. The SXA spectra at the Ti 2*p* and C 1*s* edges were measured in both TEY and TFY modes with 0.1 eV resolution at 90 and 20 incidence angles, respectively. The SXA spectra were normalized to the step edge below and far above the thresholds while the SXE spectra were normalized to the peak height. During the Ti *L* and C *K* SXE measurements, the resolutions of the beamline monochromator were 0.5, and 0.2 eV, respectively. The SXE spectra were recorded with spectrometer resolutions of 0.5 and 0.2 eV, respectively. All the measurements were performed with a base pressure lower than Torr. In order to minimize self-absorption effects [27], the angle of incidence was 20 from the surface plane during the SXE and SXA-TFY measurements. The x-ray photons were detected parallel to the polarization vector of the incoming beam in order to minimize elastic scattering. For comparison of the spectral shapes, the SXA spectra were normalized to the step edge below and far above the Ti 2*p* and C 1*s* thresholds in Fig. 3 and Fig. 4 (490 eV and 310 eV, respectively). The Ti $L_{2,3}$ and C *K* SXE spectra were normalized to the main peak heights and were plotted on a photon energy scale (bottom) and an energy scale relative to the (top) in Fig. 3 and Fig. 5.

## 3 Results

### 3.1 GI-XRD and C1s core-level XPS analysis

Results from GI-XRD and C 1*s* core-level XPS analysis are presented in Figure 1 and 2, respectively, and are in agreement with previous studies [10, 11]. When the total carbon-content increases the relative amount of a-C phase increases and the TiC grain size decreases, see summary in Table 1 and details below. In the diffractograms (Figure 1) all observed reflections can be indexed to TiC [28, 29] with a lattice parameter of 4.33-4.37 Å. The TiC grain sizes were estimated using Scherrer's equation [30, 31] to vary between 2 and 15 nm, decreasing with carbon content. This decrease in grain size represents an increase of the surface/volume ratio of the TiC grains by a factor of 7.5.

Figure 2 shows a series of C 1*s* core-level XPS spectra of the nc-TiC/a-C films. The two main peaks at 284.6 and 281.9 eV are associated with C-C and C-Ti bonds, respectivly. The smaller feature at 282.8 eV denoted C-Ti$^*$ is partly a feature of the nanocomposite, and is partly due to sputter damage which in this case should be relatively small [19]. The observed increase of the C-Ti$^*$ intensity in Fig. 2 with decreasing grain size is consistent with the possible behavior of an interface contribution. Through curve fitting analysis, the relative intensities of the three contributions were extracted and the relative amount of a-C and TiC phases estimated (see Table 1).

## 3.2 Ti 2p x-ray absorption

Figure 3 (top and middle) shows Ti $2p$ TEY- and TFY-SXA spectra following the $2p_{3/2,1/2} \rightarrow 3d4s$ dipole transitions of the nc-TiC/a-C films with different TiC grain sizes in comparison to Ti metal. The Ti $2p$ SXA intensity is proportional to the unoccupied $3d$ states and gives complementary information in the TEY and the TFY modes which depend on the incidence angle, difference in probe depths, transition processes, self-absorption and detection methods [32,33]. The SXA spectrum of pure Ti metal (black curve) has single $2p_{3/2}$ and $2p_{1/2}$ absorption peaks whereas the SXA spectra of the carbide containing materials exhibit $t_{2g}$ - $e_g$ crystal-field split double peaks due to the octahedral symmetry around the Ti site in TiC. The $e_g$ orbitals point toward the nearest C-atom sites, while the $t_{2g}$ orbitals point towards the tetrahedral holes in the fcc Ti-atom lattice. The $t_{2g}$ - $e_g$ absorption peaks at 457.6 and 459.2 eV are associated with transitions from the core shell while the $t_{2g}$ - $e_g$ peaks at 463.2 and 464.8 are associated with the core shell. Although the peak intensities vary, the Ti $2p$ crystal-field splitting (1.6 ± 0.1 eV) is the same in all spectra which is consistent with what has previously been observed in Ti $2p$ SXA spectra of crystalline bulk TiC [34]. The crystal field-splitting of TiO is known to be much larger (~ 3 eV) [35]. As observed, the intensity of the Ti $2p$ absorption increases as the grain size decreases, which shows that the number of empty Ti $3d$ states increases.

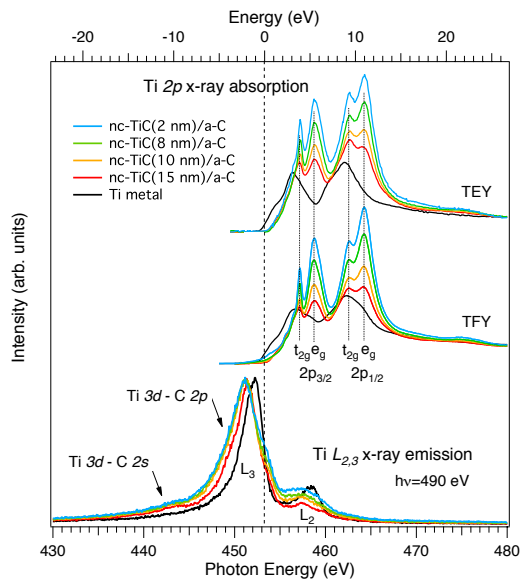

**Figure 3:** (Color Online) Top and middle: Ti $2p$ TEY- and TFY-SXA spectra of films with different grain size (in parenthesis) in comparison to bulk Ti metal. Bottom: nonresonant Ti SXE spectra of nc-TiC/a-C and Ti metal excited at 490 eV.

## 3.3 Ti x-ray emission

Figure 3 (bottom) shows Ti $L_{2,3}$ SXE spectra of nc-TiC/a-C in comparison to Ti metal, nonresonantly excited at 490 eV. Starting with the Ti $L_{2,3}$ SXE spectra of Ti metal, the main emission line is observed at -1.2 eV on the energy scale relative to the E of the component on the energy scale at the top of Fig. 3. For the Ti $L_{2,3}$ SXE spectra of nc-TiC/a-C, the main $L_3$ peak observed at -2.1 eV is due to Ti $3d$ - C $2p$ hybridization while the weak shoulder at -10 eV is due to Ti $3d$ - C $2s$ hybridization [36, 37]. The small $L_2$ emission line is found at +5.0 eV for Ti metal and +4.1 eV for the nc-TiC/a-C relative to the E of the $2p_{3/2}$ component. The low-energy shift (-0.9 eV) and the increased broadening of the $L_3$ and $L_2$ peaks for decreasing grain size are an indication of stronger Ti 3*d* - C 2*p* interaction and bonding than for the more well-defined Ti 3*d* - Ti 3*d* hybridization region in Ti metal [34]. As expected, the $t_{2g}$ - $e_g$ crystal field splitting does not appear for the occupied 3*d* valence band in the solid state consisting of overlapping bonding σ and π bands.

The larger $L_{2,3}$ spin-orbit splitting of 6.2 eV in SXE in comparison to SXA (5.6 eV) is consistent with earlier observations for TiC [34]. The trend in the $L_3/L_2$ branching ratio in transition metal compounds is a signature of the degree of ionicity in the systems [38]. For conducting systems, the / ratio is usually significantly higher than the statistical ratio 2:1 due to the additional Coster-Kronig process [27, 40]. The observed $L_3/L_2$ ratio systematically decreases as the grain size decreases, see Table I. The Ti atoms in the TiC nanocrystallites with small grain size thus appear to be more ionic and less metallic than those with larger grains. For Ti metal, the observed $L_3/L_2$ ratio of 3.74 is not directly comparable with the trend in the nanocomposite compounds.

### 3.4 C 1*s* x-ray absorption

Figure 4 shows experimental C 1*s* TEY- and TFY-SXA spectra of nc-TiC/a-C and a-C where the intensity is proportional to the unoccupied C 2*p* states. Contrary to the case of the Ti spectra, the C spectra of the nanocomposites consists of superimposed contributions from both the nc-TiC carbide and the a-C matrix. The SXA spectra were measured to identify the absorption features and peak maxima for the excitation energies of the SXE spectra presented in Fig. 5. The SXA energy region below 289 eV is known to contain $\pi^*$ resonances whereas the region above 289 eV contains $\sigma^*$ resonances [41, 42]. The increased absorption above 289 eV due to $1s \rightarrow \sigma^*$ transitions forms a broad shape resonance due to multielectron excitations.

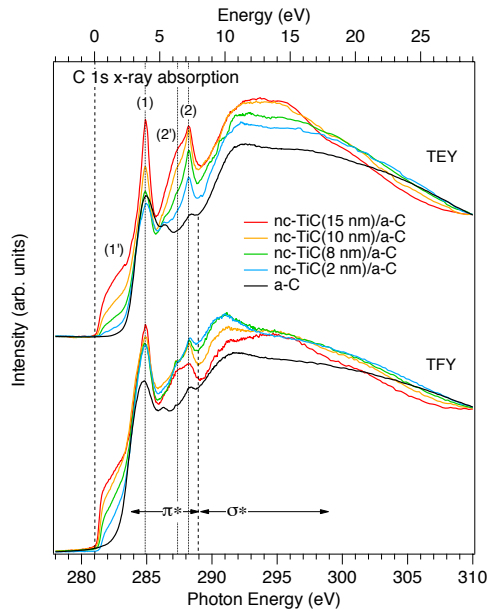

**Figure 4:** (Color Online) SXA spectra of nc-TiC/a-C films with different grain size (in parenthesis) and amorphous C with the characteristic and peak regions indicated by the horizontal arrows at the bottom. The dotted vertical lines indicate the excitation energies (1), (2') and (2) for the SXE measurements shown in Fig. 5.

The C 1*s* SXA spectra of the nc-TiC/a-C nanocomposites in Fig. 4 mainly exhibit two $\pi^*$ peaks at 284.9 eV and 288.2 eV, denoted (1) and (2), respectively. The first C 1*s* SXA peak (1) at 284.9 eV, has two contributions, partly due to C 2*p* - Ti 3*d* - $t_{2g}$ hybridization in the carbide nc-TiC nanocrystallites [43, 44] and partly due to C=C bonding contribution from the a-C matrix [45, 24]. The origin of the second $\pi^*$ peak (2) at 288.2 eV has been controversial [23, 24]. It has been suggested that peak (2) is due to either C-O bonding in an atmospherically oxidized surface layer, $sp^2$-$sp^3$ hybridization or hybridized C $2p$ - transition metal $3d$ - $4sp$ states. As observed in Fig. 4, the intensity of peak (2) is higher in the surface-sensitive TEY spectra than in the more bulk-sensitive TFY spectra [22] and the relative intensity of peak (2) is known to largely depend on the sample structure and composition by different deposition methods [20, 21]. However, the peak intensity in TFY also depends on the incidence angle and is reduced due to self-absorption effects at normal incidence.

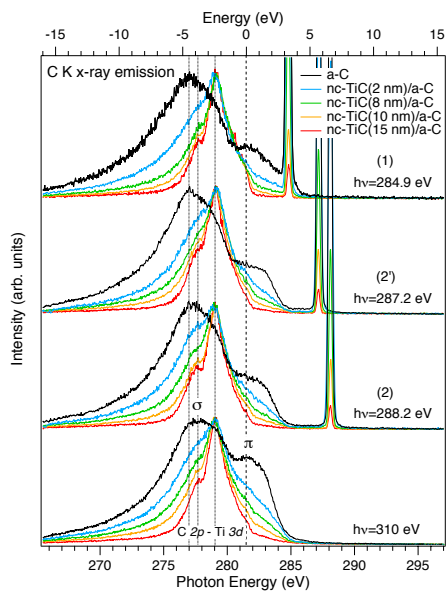

**Figure 5:** (Color Online) Resonant and nonresonant C $K$ SXE spectra of nc-TiC/a-C films with different grain size (in parenthesis) and amorphous C excited at peak (1) at 284.9, the shoulder (2') at 287.2 eV and peak (2) at 288.2 eV in the SXA spectra and nonresonant at 310 eV. The occupied σ and π bands of a-C are indicated at the bottom.

Peak (1) has a broad and pronounced low-energy shoulder (1') below the main peak at 281-283 eV that is solely due to the carbide contribution. The low energy shoulder reflects the gradual carbide formation for increasing grain size as seen in Table I. Peak (2) also has a low-energy shoulder (2') at 286-288 eV. The observed intensity quenching of the shoulders (1') and (2') for decreasing grain size implies a depletion of unoccupied C $2p$ states as the carbide contribution decreases. Although part of peak (2) in the surface sensitive TEY-XAS spectra can be attributed to C-O bonding due to atmospheric oxidation at the surface, the more bulk sensitive TFY spectra show that it is also due to C $2p$ - Ti $3d$ - $e_g$ hybridization in TiC with addition of the superimposed a-C contribution. The apparent $t_{2g}$ - $e_g$ splitting originating from the unoccupied Ti $3d$ orbitals is here indirectly observed in the C $1s$ SXA spectra, but is significantly wider (2-3 eV) than for the Ti $2p$ SXA spectra in Fig. 3 (1.6 eV). The intensity around 291 eV systematically increases as the grain size decreases with increasing amount of a-C. The a-C spectrum has a well-defined shape-resonance structure around 292 eV observed both in TEY and TFY. Note that the intensity trend is largely opposite at 282 eV in comparison to at the a-C shape-resonance at 292 eV.

## 3.5  C *K* x-ray emission

Figure 5 shows C $K$ SXE spectra excited at peak (1) in the SXA spectra at 284.9 eV, at the shoulder (2') at 287.2 eV and at peak (2) at 288.2 eV (resonant) and 310 eV (nonresonant) photon energies, probing the occupied C 2*p* states of the valence bands. As in the case of the C 1*s* SXA spectra, the C *K* SXE spectra of the nanocomposites represent a superposition of two contributions from the nc-TiC carbide and the surrounding a-C matrix. As the C 2*p* intensity is the largest in the case of the smallest TiC grain size, the intensity trend is now opposite from the SXA spectra shown in Fig. 4. As can be seen in Fig. 5, most intensity in the upper valence band is observed for the C *K* SXE spectra of a-C (black curve). The main peak corresponding to the occupied C 2*p* σ band is observed at 277 eV photon energy with a broad high-energy band at 281.5 eV with π character. The C *K* SXE spectra of a-C exhibit a similar spectral shape as $sp^2$ hybridized graphite [46]. However, the intensity of the π band at 281.5 eV is higher for a-C than in graphite, indicating additional influence of hybridized C at the top of the valence band as in the case of hybridized diamond [46]. In comparison to C *K* SXE spectra of graphite and diamond, the spectral shapes of a-C and the nanocomposites show less excitation energy dependence. For a-C, the σ/π intensity ratio remains constant (2.4) for the resonant photon energies while it decreases (1.4) for the non-resonant excitation at 310 eV.

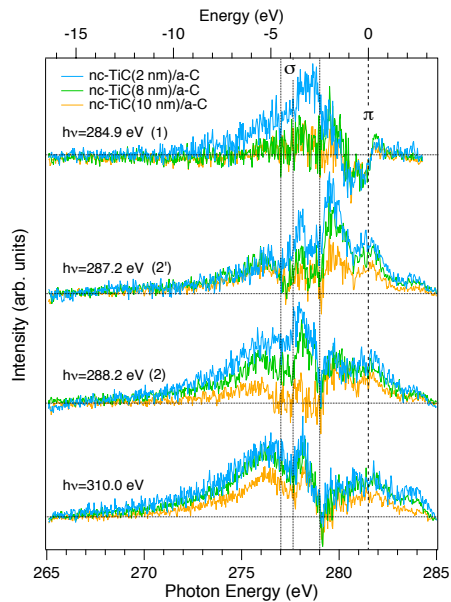

**Figure 6:** (Color Online) Difference plots of resonant and nonresonant C *K* SXE spectra from Fig. 5 of three nc-TiC/a-C films with different grain size (in parenthesis) obtained by subtraction from a superposition of the spectra with the 15 nm grain size and a-C excited at peak (1) at 284.9, at the shoulder (2') at 287.2 and peak (2) at 288.2 eV in the SXA spectra and nonresonant at 310 eV.

The details of the spectral SXE features of the nanocomposites depend not only on the overlapping relative spectral contributions of the π and σ bands of the a-C matrix but also on the addition of the superimposed contributions from nc-TiC and possible interface states. Contrary to the case of a-C, the C *K* SXE spectra of the nanocrystallites with 15 nm size (red curve), have the characteristic spectral shape of TiC [34]. It has a sharp main peak at -2.5 eV and a low-energy shoulder at -3.8 eV below the E of TiC at the energy scale at the top of Fig. 5. The sharp -2.5 eV peak is characteristic of the strong covalent Ti 3*d* - C 2*p* bonding in TiC [39, 34]. Note that the -3.8 eV low-energy shoulder is sharpest for the spectra excited at 284.9 eV and 288.2 eV corresponding to the main SXA peaks (1) and (2) in Fig. 4. A high-energy shoulder between the σ and the π bands of the nanocomposites at -1.0, is most developed for the spectra excited at 284.9 eV (1), indicating contribution of C 2*p* - Ti 3*d* - $t_{2g}$ hybridization in this energy region (0 to -2 eV). For the spectra excited at the other excitation energies, the high-energy shoulder is absent. This implies a different type of bonding symmetry for excitation at peak (1) than at peak (2).

In Fig. 5, the SXE spectra of the nanocomposites consist of a sum of the C contributions from the nc-TiC carbide and the surrounding a-C matrix. In addition, there is a possible contribution from an interface state at the surfaces of the nc-TiC crystallites as suggested by the C-Ti* peak in XPS (Fig. 2). A way to resolve the interface contribution is to make difference plots based on the C $K$ SXE spectra in Fig. 5. This was done by a weighted superposition of the spectra from the a-C sample and the 15 nm sized nanocomposite, which has a very similar spectral shape as bulk TiC. The same weight factors were used for all four excitation energies and the results are shown in Fig. 6. For the 2 nm sample, weight factors of 0.60 and 0.40 were used. For the 8 nm sample, weight factor of 0.82 and 0.18 were used. For the 2 nm sample, weight factors of 0.93 and 0.07 were used. The intensity and integrated area of the difference spectra increases as the grain size decreases. This would be consistent with an interface state or component, which should grow for smaller grain sizes. The shape of the difference spectra largely depends on the excitation energy. For the spectra excited at peak (1) at 284.9 eV, the intensity difference is negative in the $\pi$ energy region 280-282 eV just below the E of TiC. The negative intensity is a signature of broken Ti-C bonds with orbitals of $t_{2g}$ symmetry in TiC. This observation is consistent with theoretical studies of C adsorbed on TiC [47]. The difference spectrum excited at the shoulder (2') at 287.2 eV has a main peak at -1.8 eV (279.7 eV), which is much lower for the other excitation energies. This is an indication of an additional spectral component in the SXA spectrum at 287.2 eV (2') and probably a result of additional hybridization with the Ti 3$d$ orbitals of $e_g$ symmetry which hybridize with the C 2$p$ states at this excitation energy in the C 1$s$ SXA spectra.

**Table 1:** The first column gives the estimated grain size of the TiC nanocrystallites from XRD and the second column gives the total C content of the samples. The third and fourth column gives the relative amount of bonds from peak-fitting of the XPS spectra in the C 1$s$ region in Fig. 2. The last column shows the $L_3/L_2$ ratio in the Ti SXE spectra in Fig. 3.

| Grain size | C tot | C-Ti + C-Ti* | C-C | $L_3/L_2$ |
|---|---|---|---|---|
| 15 nm | 35% | 93% | 7% | 6.56 |
| 10 nm | 42% | 76% | 24% | 5.12 |
| 8 nm | 51% | 60% | 40% | 4.72 |
| 2 nm | 65% | 34% | 66% | 4.08 |

# 4 Discussion

In the SXA and SXE spectra of Ti and C in Figs. 3-6, there are several interesting observations. For Ti in the nc-TiC carbide phase in Fig. 3, the number of unoccupied Ti 3$d$ states increase with decreasing grain size both in TEY and TFY. The relative intensities between the $t_{2g}$ - $e_g$ ligand field peaks observed in Ti 2$p$ SXA show that the unoccupied 3$d$ states of the Ti atoms in the smallest nanocrystalline TiC grains are most affected by charge-transfer from the Ti 3$d$ to the C 2$p$ orbitals. This charge-transfer may occur within the TiC nanocrystals but more likely across the interface to the surrounding a-C phase. Comparing the Ti 2$p$ SXA and SXE spectra in Fig. 3 of the nc-TiC/a-C nanocomposites with those of Ti metal, the spectral intensities are largely affected by the grain size of the TiC nanocrystallites. At the same time as the number of unoccupied Ti 3$d$ states increase, we observe that the number of occupied Ti 3$d$ states decreases as the grain size decreases. This is consistent with an increased charge-transfer from the Ti 3*d* to the C 2*p* orbitals as the grain size decreases. The increased charge-transfer is also reflected in the / branching ratio which shows that the ionicity increases as the grain size decreases.

For the superimposed C contributions from the nc-TiC phase and the a-C matrix in Figs. 4 and 5, the number of unoccupied states decreases and the number of occupied states increases as the grain size decreases. In C 1*s* SXA, the intensities of two pronounced peaks due to C 2*p* - Ti 3*d* - $t_{2g}$ and C 2*p* - Ti 3*d* - $e_g$ hybridization show strong variation with grain size. The origin of the absorption peak (2) between the and C 1*s* absorption resonances has been controversial [22, 23, 24] and has been assigned to either C-O bonding, $sp^2$-$sp^3$ hybridization or hybridized C 2*p* - transition metal 3*d* - 4*sp* states at the interface. Although part of peak (2) in surface sensitive TEY-XAS can be attributed to C-O bonding due to atmospheric oxidation at the surface, the more bulk sensitive TFY spectra show that it is also due to C 2*p* - Ti 3*d* - $e_g$ hybridization in TiC with addition of the superimposed a-C contribution. The apparent $t_{2g}$-$e_g$ splitting originating from the unoccupied Ti 3*d* bands is here indirectly observed in the C 1*s* SXA spectra, but is significantly broader (2-3 eV) than for the Ti 2*p* SXA spectra in Fig. 3 (1.6 eV). For the C *K* emission spectra, an increased number of occupied states is observed as the C content is increased for the smaller TiC crystallites. Note that this trend is opposite from the case of the Ti SXA/SXE spectra in Fig. 3 and implies an increased charge-transfer from Ti to C as the grain size decreases. The largest C *K* emission intensity is observed for amorphous C which contains the most occupied C 2*p* electronic states with distinguishable σ and π bands.

From the spectral features in the SXE spectra in Figs 3 and 5, two main types of bonds were also identified; the strong Ti 3*d* - C 2*p* carbide bonding in the TiC phase and the weaker C 2*p* - C 2*p* bonding in the a-C phase. The Ti 3*d* - C 2*p* covalent bond region is concentrated to a specific energy region -2.5 eV below E while the C 2*p* - C 2*p* hybridization generally occurs in a much wider energy region with overlapping but distinguishable σ and π bands. From the C *K* SXE difference spectra of weighted superpositions in Fig. 6, the intensity and the integrated area increases as the grain size decreases. The SXE intensity is also largely quenched just below E for the excitation energy at the main peak (1) and additional SXE intensity occurs between the occupied σ and π C bands for excitation at the C 1*s* SXA shoulder (2'). These observations are correlated with a possible interface state or phase contribution as previously observed as a feature denoted C-Ti in the C 1*s* XPS spectra in Fig. 2. When the grain size decreases from 15 nm to 2 nm, it implies that the interface/bulk ratio increases for the nc-TiC carbide phase by a factor of 7.5. Although an interface component is not directly observed in the C *K* SXE spectra, it is identified in the difference spectra in Fig. 6. The C *K* SXE difference spectra reveal a spectral component with more broken Ti-C bonds of orbitals with $t_{2g}$ character and additional bonds with orbitals with $e_g$ character for increasing interface/bulk ratio as the nc-TiC grains become smaller.

## 5 Conclusions

The electronic structure, chemical bonding and interface component of nanocomposites of TiC crystallites embedded in a matrix of amorphous carbon has been investigated using x-ray absorption and x-ray emission spectroscopy. A strong intensity dependence of the TiC grain size is observed. The trend in the $L_3/L_2$ branching ratio of Ti $L_{2,3}$ x-ray emission indicates an increased ionicity of the Ti atoms as the grain size decreases. This is caused by increased charge-transfer from the Ti 3*d* to the C 2*p* states as the grain size decreases. Analysis of C *K* x-ray emission difference spectra by weighted spectral superposition reveals increased spectral intensity as the grains size decreases. This is consistent with an additional component at the TiC/amorphous carbon interface as previously suggested by C 1*s* photoelectron spectra. It is found that the additional intensity involves more Ti 3*d* orbitals of $e_g$ symmetry than $t_{2g}$ symmetry. This suggests a distinctly different bonding at the surface of the TiC nanocrystallites embedded in amorphous carbon, where an increased charge-transfer from the Ti 3*d* $e_g$ orbitals at the interface to the C 2*p* orbitals in the amorphous C phase causes an increased ionicity of the entire TiC nanocrystallites. An interface bonding in the nanocomposites may affect the physical properties such as electrical conductivity and possible hardness or elasticity. Since understanding the interfacial bonding in a nanocomposite is essential to enable the design of the materials properties, this will be the subject of further spectroscopic and theoretical studies.

# 6 Acknowledgements

We would like to thank the staff at MAX-lab for experimental support. This work was supported by the Swedish Research Council, the Göran Gustafsson Foundation, the Swedish Strategic Research Foundation (SSF), Strategic Materials Reseach Center on Materials Science for Nanoscale Surface Engineering (MSE).